\documentclass[fleqn,usenatbib]{mnras}

\usepackage[T1]{fontenc}

\DeclareRobustCommand{\VAN}[3]{#2}
\let\VANthebibliography\thebibliography
\def\thebibliography{\DeclareRobustCommand{\VAN}[3]{##3}\VANthebibliography}

\newcommand{\mr}{\mathrm}
\newcommand{\rco}{r_\mr{co}}
\newcommand{\pms}{P_{-3}}
\newcommand{\Bpsr}{B_{10}}
\newcommand{\Mpsr}{M_\mr{psr}}
\newcommand{\sm}{M_\mr{sun}}

\usepackage{graphicx}	
\usepackage{amsmath}	
\usepackage{amssymb}	

\usepackage{newtxtext,newtxmath}
\title[rotational-powered FRBs]{FRBs from rapid spindown neutron stars}

\author[Li and Pen]{
Dongzi Li$^{1,2}$\thanks{E-mail: dongzili@princeton.edu}
and Ue-Li Pen$^{3,4,5,6,7}$
\\
$^{1}$TAPIR, California Institute of Technology, Pasadena, CA 91125, USA \\
$^{2}$ Department of Astrophysical Sciences, Princeton University, Princeton, NJ 08544, USA\\
$^{3}$Institute of Astronomy and Astrophysics, Academia Sinica, Astronomy-Mathematics Building, No. 1, Section 4,
Roosevelt Road, Taipei 10617, Taiwan \\
$^{4}$Canadian Institute for Theoretical Astrophysics, University of Toronto, 60 St. George Street, Toronto, ON M5S 3H8, Canada\\
$^{5}$Perimeter Institute for Theoretical Physics, 31 Caroline St. North, Waterloo, ON, Canada N2L 2Y5\\
$^{6}$Canadian Institute for Advanced Research, CIFAR program in Gravitation and Cosmology\\
$^{7}$Dunlap Institute for Astronomy \& Astrophysics, University of Toronto, AB 120-50 St. George Street, Toronto, ON M5S 3H4, Canada
}

\date{Accepted XXX. Received YYY; in original form ZZZ}

\pubyear{2023}

\begin{document}
\label{firstpage}
\pagerange{\pageref{firstpage}--\pageref{lastpage}}
\maketitle

\begin{abstract}

A fast radio burst (FRB) localized to a globular cluster (GC)
challenges FRB models involving ordinary young magnetars. In this
paper, we examine the rapid spindown millisecond neutron star (NS) scenario, which favours
the dynamic environment in GCs. Fast spindown
corresponds to a larger 
magnetic field than regular millisecond pulsars, which empirically
favours giant pulse (GP) emission. 
The kinetic energy in millisecond NSs can readily exceed the magnetic
energy in magnetars. 
The high inferred isotropic luminosity of most FRBs is challenging to
explain in spin-down powered pulsars.
A recent observation of
a GP from the Crab pulsar, on the other hand, suggests highly Doppler-beamed
emission, making the required energy orders of magnitude smaller than
estimated with isotropic assumptions. 
Considering this strong beaming effect, GPs from a recycled
pulsar with a modest magnetic field could explain the energetics and burst rates for a wide
range of FRBs. The short life span accounts for a paucity of bright
FRBs in the Milky Way neighbourhood. We point out that tidal
disruption spin-up from a main sequence star can provide sufficient accretion rate to recycle a NS with mild magnetic field. It can also explain the observed source density and the spatial offset in the GC for FRB 20200120E. Frequency variation in the scattering tail for some of the brightest FRBs is expected in this scenario. 

\end{abstract}




\section{Introduction}
Fast radio bursts (FRBs) are enigmatic radio bursts with durations
ranging from microseconds ($\mu$s) to milliseconds (ms). Over the past
decade, substantial advances have been made in constraining the nature
of FRBs. The identification of repeating FRBs
\citep[e.g.,][]{2014Spitler,2016Spitler,CHIME2019c,Fonseca+2020}
suggests that at least a subset of the FRB population originates from
non-cataclysmic events. The detection of exceptionally intense radio
bursts from a galactic magnetar \citep{20SGRCHIME,20SGRSTARE2} has
positioned magnetars as a popular candidate progenitor for
FRBs. However, it is unclear whether all FRBs are produced
by ordinary magnetars.
Evidence is building for a diverse population of FRBs.   
While magnetars are young and closely related to the star-formation in
the past $<10^5$ year, here is a growing variety of FRB host galaxies
including in elliptical galaxies with little star formation. 
Moreover, the star-formation history among host galaxies exhibits a
broad delay-time distribution, covering a range from approximately 100
million years (Myr) to around 10 billion years (Gyr)\citep{2023Law},
which are both much longer than the magnetar life time. Two of the
most active repeaters, FRB 20180916A and FRB 20121102B exhibit
long-term periodicity \citep{CHIME2020b, Rajwade2020}, while no known
magnetars exhibit similar properties.  
The definite evidence supporting an alternative formation channel is
provided by the localization of FRB 20200120E
\citep{Majid2021,Bhardwaj2021} to an old ($t\gtrsim 10\,$Gyr) globular
cluster (GC) in M81 \citep{Kirsten2022}. These old GCs have not
experienced massive star formation for billions of years, effectively
ruling out the possibility of a young magnetar born through a massive
stellar collapse 
Alternatively, \citet{Kirsten2022,Kremer2021,Lu2022} proposed that
magnetars formed in accretion-induced collapse
\citep[e.g.,][]{Tauris2013} or via massive white dwarf binary merger
may provide a formation mechanism for this repeater.  
Nevertheless, in simulations, there is an ongoing debate regarding
whether the mass can be retained to form magnetars or if it will be
ejected during the collision process. 


In this paper we examine the scenario of giant pulses from rapid spindown neutron stars
as a mechanism for FRBs. In section
\ref{sec:spindown} We show that it could satisfy the key observation constraints: rate, spectral luminosity, and total emitted energy.  In section \ref{sec:spinup} we consider a
generation mechanism due to tidal disruption spinup. In section \ref{sec:rate}, we discuss the rate and spatial distribution regarding the scenario. 
We summarize and discuss the proposed observation tests in
section \ref{sec:summary}.


\begin{figure}
	\includegraphics[width=\columnwidth]{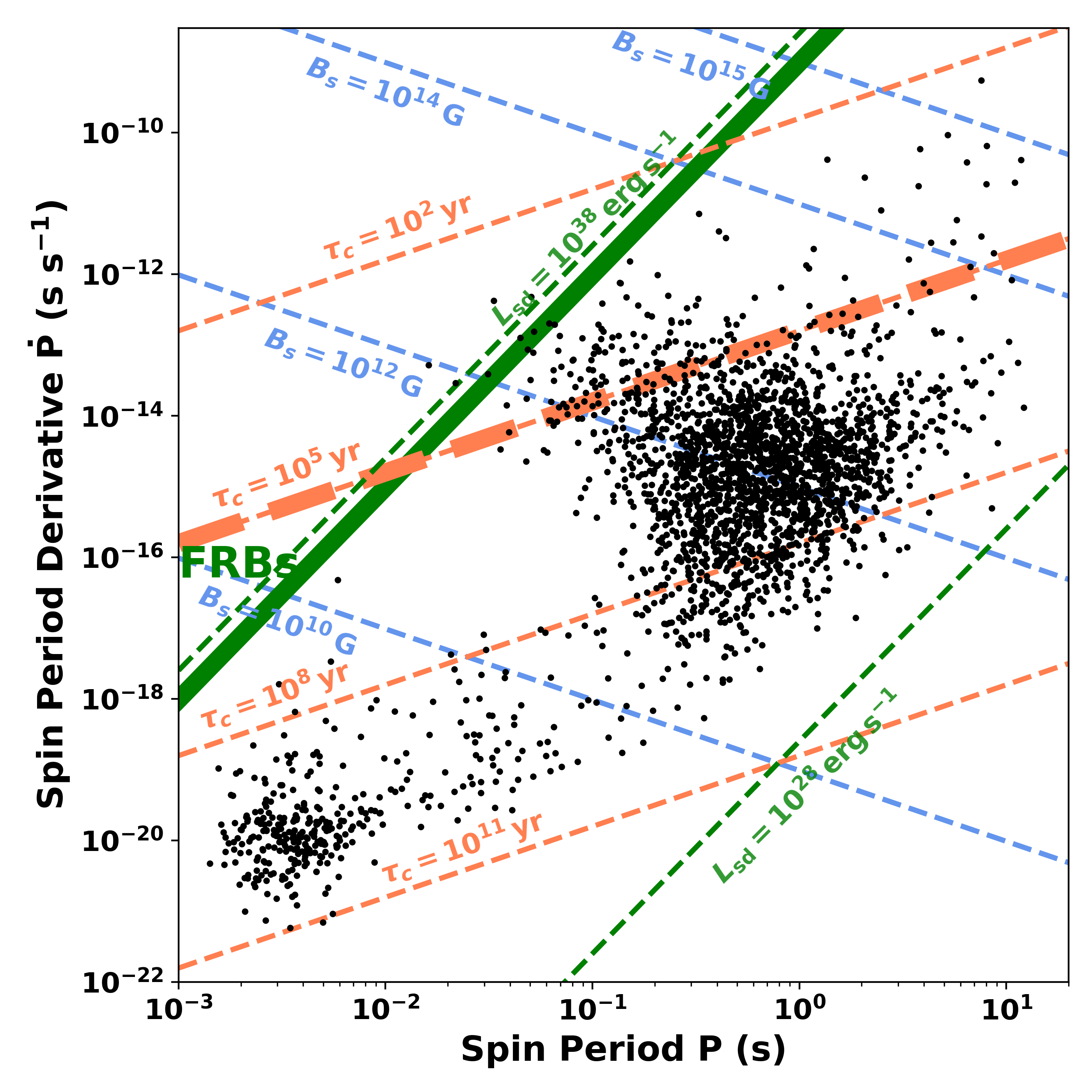}
    \caption{P-$\dot{P}$ diagram of rapid spindown scenario.  The
      thick orange dashed line is the spin-up limit, which is similar to the
      abundance limit, while the thick green solid
      line represents the observed mean energy at 1\% radiation
      efficiency. Dots represent known pulsars\citep{2016yCat....102034M}.
}
    \label{fig:ppdot}
\end{figure} 

\section{Energetics}
\label{sec:spindown}

With the notable exception of magnetars, most pulsars are powered by
their spindown energy. The rapid rotation of the neutron star leads to
magnetic dipole radiation:  
\begin{equation}
    P_\mr{rad}=\frac{2}{3c^3}(\mu\sin\alpha)^2\big(\frac{2\pi}{P}\big)^4
    \sim 10^{39}\mr{erg}/s \, \Bpsr^2 \pms ^{-4}
\end{equation}
where $\Bpsr$ is the surface magnetic field in units of $10^{10}$ G, 
$\mu$ is the magnetic dipole moment and $\pms$ is the spin period in ms. 

Typically radio emission only accounts for a small fraction of the
spindown energy, with most of the energy dissipated in a pair plasma
wind, some of which is visible by the reconnection processes at the
light cylinder.  We will discuss below how this might change for
giant pulse emission in the presense of strong magnetic fields at the
light cylinder. The peak flux of a radio burst can be estimated with 
\begin{equation}
S_\mr{pk}=P_\mr{rad} f_r \Delta\nu^{-1}\Omega^{-1}d^{-2} N^{-1}    
\end{equation}
where $f_r$ is the fraction of energy emitted in the radio frequency during the burst; $\Delta \nu$ is the bandwidth of the burst; $\Omega$ is the solid angle of the emission; $d$ is the distance to the Earth and $N$ is the number of bursts emitted simultaneously. 

The isotropic equivalent spectral luminosity for a single burst would be: 
\begin{align}
    L_\nu&=S_\mr{pk}4\pi d^2 =P_\mr{rad} f_r \Delta \nu^{-1} N^{-1} \frac{4\pi}{\Omega}\nonumber\\
    &= 10^{36}\mr{erg/s/Hz} \, \Bpsr^2 \pms^{-4} f_{r,-3} \Delta\nu_9^{-1} N^{-1} \Omega^{-1}_{-8} 
    \label{eq:Lnu}
\end{align}
Here we assume $B=10^{10} G$, millisecond period, $\Delta\nu=1$ GHz and radio efficiency $f_r$ is
$10^{-3}$, which is observed in the case of the galactic magnetar SGR 1935+2154 FRB-like burst. 

The GP scenario has been discussed previously by \citet{CordesChatterjee2019}. 
For a wide angle emission, ${\Omega\sim 4\pi}$, a milisecond neutron star with an ordinary magnetic field will not produce enough energy for an FRB, which all have spectral luminosity $L_\nu>10^{27} \mr{erg/s/Hz}$. 
However, coherent radio radiation is expected to be strongly beamed. 
For radio pulsars, the bulk motion of emitting plasma is expected to
have Lorentz factors greater than $10^2$\citep{2021MNRAS.500.4530M}.
Recent studies of pulsar scintillation suggest highly
relativistic motion in a bulk
coherent flow in the emission region of the Crab giant pulses. The measured Lorentz factors $\gamma \gtrsim
10^4$\citep{21Bij,2022arXiv221105209L} leading to beaming of 
giant pulses into a tiny solid angle $\Omega\sim 1/\gamma^2$.   We
will assume the beaming directions to be random.
The measured high Lorentz factors can help explain the radio emission with free electron
lasers\citep{2021ApJ...922..166L} or moving
mirrors\citep{2022MNRAS.515.5682Y}.
With $\gamma\sim 10^4$, we have forward gain of $4\pi/\Omega\sim 10^7$. 
The nature of coherent emission, especially that of giant pulses, is
still not well understood.  Giant pulses are preferentially observed
to occur in pulsars with strong magnetic fields at the light
cylinder $B_{\rm LC}>10^5$G\citep{2006ApJ...640..941K}.
Extrapolating, rapid spindown pulsars might dominate their spindown
energy through the emission of giant pulses. 
In this case, the spindown energy is able to account for most of the FRBs with known redshift. 

The observed burst rate can be parametrized: 
\begin{align}
    R_\mr{obs}=RN \frac{\Omega}{4\pi}
    \label{eq:rate}\end{align}
where $R$ is the rate of the FRB bursts from this source in all directions. 
The separation of the bursts has to be greater than the burst duration $W$ for them to be separated in time. Therefore, the number of bursts emitted at the same time $N\geq R W$. Together with Eq.~\ref{eq:rate}, we have
\begin{align}
   N\geq \sqrt{R_\mr{obs}W 4\pi/\Omega}
   \label{eq:n}
\end{align}

The total energy of the observed bursts accumulated over time $\Delta T$:
\begin{align}
    E_\mr{T}=f_r P_\mr{rad} \Delta T
    \label{eq:E}
\end{align}
The total fluence will be: 
\begin{align}
    F_\mr{T}=f_r P_\mr{rad} \Delta T/4\pi d^2
\end{align}
After accumulating over time scales much longer than the burst
separations, the total fluence does not depend on the solid angle.  

For FRB 20200120E detected in the GC of M81, the observed isotropic equivalent spectrum luminosity $L_\nu\sim10^{27} - 10^{28}$ erg/Hz/s. As shown in Equation~\ref{eq:Lnu}, an ordinary recycled pulsar with a field strength of $10^8$~G and a period of 10~ms should in principle have enough spin down energy to produce the bursts given a beaming angle similar to the Crab pulsar. 

For FRB 20201124A, energetic bursts of a few times $10^{34}$ erg/s/Hz have been detected with more than 2000 hr observation with 25-32m telescopes. As seen from Eq~\ref{eq:Lnu}, the spindown energy
from a milisecond neutron star with $10^{10}$ G field is enough to
produce the bursts in the highly beamed scenario. 

For FRB 20121102A, the observed isotropic equivalent spectrum luminosity ranges
$S_\mr{pk}\sim 10^{30}-10^{34}$ erg/Hz/s. The peak burst rate can reach 122
hr$^{-1}$ with the FAST observation, with the majority of the burst
$S_\mr{pk}\sim 10^{31}$. Assuming a highly beamed emission with
$\omega\sim 10^{-8}$ as discussed before, we expect N$\geq 300$ for
the faint bursts following Eq~\ref{eq:n}.  N can be 1 for the
energetic bursts. As seen from Eq~\ref{eq:Lnu}, the spindown energy
from a milisecond neutron star with $10^{10}$ G field is enough to
produce the bursts in the highly beamed scenario.  The total isotropic
energy emitted by the 1652 bursts detected in 59.5 hour spanning 47
days is $3.4\times10^{41}$ erg, corresponding to a mean luminosity of
$\sim 10^{36}$ erg/s.  Following Eq~\ref{eq:E}, the total
energy that can be provided by spin down of a milisecond pulsar in
59.5~h is $10^{45} \mr{erg} f_r B_{10}^2 \pms ^{-4}$ which is enough
to produce the bursts assuming $f_rB_{10}\gtrsim 10^{-3}$ during the
radio peak activity time.

\section{Spinup}
\label{sec:spinup}
As discussed in the last section, orders of magnitude larger magnetic field than the typically observed millisecond pulsars is required to explain the spectral luminosity of most of the FRBs. 
We discuss the physical mechanism to generate this kind of short lived millisecond pulsars in
globular clusters.  For simplicity, we consider a neutron star in the
center of a GC tidally partially disrupting a main sequence star in a
close encounter.   \citet{1996JKAS...29...19L, 2022ApJ...934L...1K} examined
numerical simulations of such disruptions, with the goal to explain
isolated recycled pulsars in GCs.  Depending on impact parameter, a
fraction of the close encounter star can be disrupted, with most of
the stripped off material bound to the neutron star.  A significant portion of
this material will ultimately accrete onto the neutron star, resulting
in an increase of angular momentum. 
In an accretion disk, angular momentum transfer occurs by inner mass
transferring angular momentum to outer mass.  At the Alven radius $r_A$, the point of pressure balance
between the accretion disk and magnetic field,
mass flows in directly, transferring orbital angular momentum to
the central object.  While the details in the accretion needs dedicated simulations, we discuss few basic limits here. 

For neutron star with large magnetic field, the required accretion rate is large. 
Spinup only occurs
while the Keplerian speed is higher than the co-rotation speed at
$r_A$.  
For materials accreting onto a magnetized neutron star, the magnetic
energy density balances the kinetic energy density $B^2/8\pi=\rho
v^2/2$ at the Alfven radius $r_A$. Assuming the matter moves in
spherical radial free fall, $\rho=\dot{M}/4\pi v r^2$ and
$v=\sqrt{2G\Mpsr/r}$.  The relevant magnetic field at the Alfven radius is the dipole component $B=\mu/r^3$. Then the Alfven radius can be estimated with:  
\begin{align}
    r_A=\big(\frac{\mu^4}{2G\Mpsr\dot{M}^2}\big)^{1/7} =25\mr{km}\,\Bpsr^{4/7}\dot{M}_{-5}^{-2/7}
    \label{eq:rA}
\end{align}
where $\mu=BR^3$ is the magnetic dipole moment, $R$ is the neutron
star radius. $\dot{M}$ is in unit of $\sm/\mr{yr}$.  

To spin-up the neutron star, the Alfven radius has to be smaller
than the co-rotation radius $\rco=(G\Mpsr P^2/4\pi^2)^{1/3}=20km P_{-3}^{2/3}$. At $\rco$
the Keplerian angular velocity equals the spin angular velocity. If
material outside $\rco$ interacts with the star via the magnetic
field, it will be spun up and repelled, slowing down the neutron
star. For millisecond pulsar, the co-rotation radius is close to the neutron star radius. Therefore, a minimum accretion rate is required for the spin-up to happen: 
\begin{align}
    \dot{M}_\mr{min}= \frac{1}{\sqrt{2}} \mu^2 \big(\frac{2\pi}{P}\big)^{7/3}(G\Mpsr)^{-5/3}
    =10^{-5}\sm/\mr{yr}\,\Bpsr^{2} \pms^{-7/3} 
    \label{eq:Mdot}
\end{align}

After the accretion, the angular momentum of the neutron star will be
increased by $\Delta J=\Delta M vR=\Delta M \sqrt{2G\Mpsr R}$  
and therefore, it will be spun up to 
$P=2\pi I/\Delta J$. 
For a neutron star to reach millisecond period, one needs
\begin{align}
    \Delta M=I\frac{2\pi}{P} (2G\Mpsr )^{-1/2} \mr{max}(R,r_A)^{-1/2}
    \approx 0.16 \sm \, \pms^{-1}
    \label{eq:dM}
\end{align}
And all the material has to be accreted within 
\begin{align}
	T_\mr{max}=\frac{\Delta M}{\dot{M}_\mr{min}}=10^4\mr{yr}\, B_{10}^{-2}P_{-3}^{4/3} 
\end{align}
The required $\dot{M}$ is substantial
to create a milisecond pulsar with moderate magnetic field. 
A significant fraction of the companion
star has to be accreted to the NS within less than $10^4$ year for
$B=10^{10}$ G, and less than a year for an ordinary pulsar of
$B=10^{12}$ G, and within an hour for a magnetar of $10^{14}$ G.  

The upper limit of the accretion rate can be estimated with 
the shortest possible time scale -- the tidal disruption time. The neutron star has a close encounter with the main sequence star at
a timescale $T\approx d/v_r$. The impact distance $d$ can be approximated with the radius of the main sequence star $R_\mr{star}$ as the other relevant scales $r_A$ and $R$ are both much smaller.  $v_r=\sqrt{2GM_\mr{star}/R_\mr{star}}$ is the
relative velocity of the two object.  
Therefore 
\begin{align}
T_\mr{min}\approx d/v_r=\sqrt{2R_\mr{star}^3/GM_\mr{star}}=0.6\mr{hr} \, M_\mr{star}^{-1/2}R_\mr{star}^{3/2}	
\end{align}
where $M_\mr{star}$ and $R_\mr{star}$ are in the unit of solar mass and solar radius respectively. Hence the maximum accretion rate will be:  
\begin{align}
    \dot{M}_\mr{max}&=\frac{\Delta M}{T_\mr{min}} =I\frac{\pi}{P} (\frac{M_\mr{star}}{\Mpsr})^{1/2} R_\mr{star}^{-3/2} \mr{max}(R,r_A)^{-1/2}\\
    &=2\times10^{4} \sm/\mr{yr} \,
    M_\mr{star}^{1/2}R_\mr{star}^{-3/2} \pms^{-1} 
    \label{eq:MdotEncounter}
\end{align}
To recycle a neutron star to millisecond $\dot{M}_\mr{max}\gtrsim\dot{M}_\mr{min}$, hence, 
\begin{align}
	B_{10}P_{-3}^{-2/3}<10^4   M_\mr{star}^{1/4}R_\mr{star}^{-3/4} 
\end{align}
Therefore, it is not possible to recycle magnetars with $B\gtrsim 10^{14}$ to milisecond. Actually, the accretion timescale can be much longer than the tidal dynamical time scale. According to \citet{2022ApJ...934L...1K}, the viscous accretion time is approximately 1 to 5 days. In this case, magnetars with $B\gtrsim 10^{13}$ G are difficult to be recycled to millisecond. However, for most of the ordinary neutron stars with magnetic field $10^{12}$ G or lower, it is possible to recycle them to millisecond in the tidal disruption events. 

Another challenge for the large accretion rate is the large radiation pressure. 
As the mass accretes, it converts the gravitational potential 
energy into thermal energy and exerts a radiation force
$F_\mr{rad}=\kappa m L/4\pi R^2c$ which opposes the gravity force
$F_\mr{grav}=GMm/R^2$. Here $\kappa$ is the opacity, which is defined
as the cross-section per unit-mass. For ionized hydrogen,
$\kappa=\sigma_T/m_p$, where $\sigma_T$ is the Thomson cross section
and $m_p$ is the mass of the proton. $m$ is the mass of the infalling
material; $c$ is the speed of light. And the luminosity $L$ can be
estimated as a fraction $\epsilon$ of the total energy of the accreted
material $L=\epsilon\dot{M}c^2$. This gives the Eddington accretion
rate: 
\begin{align}
    \dot{M}_\mr{EDD}=\frac{4\pi GM m_p}{\epsilon c\sigma_T}\approx 10^{-9} \sm/\mr{yr} \,
    \epsilon^{-1}
\end{align}
Even for a relatively low magnetic field pulsar $B=10^{10}$ G, the
required accretion rate is still orders of magnitude higher than the
Eddington rate. This kind of ``hypercritical" accretion has been
considered in various fields, including accretion during the
supernovae and the common envelope evolution. It can be featured in
several ways. With sufficiently high mass inflow rate
(e.g. $\dot{M}_\mr{cr}\approx10^{-4} \sm \mr{yr}^{-1}$
\citealt{1996Fryer}), the photons can be trapped and advected inwards,
where the neutrino cooling plays a role and allows for hypercritical
accretion\citep{1993Chevalier,1996Chevalier}. Apart from this, the
accretion and radiation do not occur in the same
direction. The inward mass flow predominately occurs in the plane of
the disk while the accretion energy can flow out low-density polar
regions\citep{2002apa..book.....F}.  
Observationally, a high-mass X-ray
binary has been observed to accrete at 100 times Eddington accretion
rate\cite{2014Natur.514..202B}.  


In summary, more than 10\% of the solar mass is needed to spin-up a neutron star of $10^{10}$ -- $10^{13}$ G magnetic field to millisecond. The required large accretion rate is plausible to be achieved in the partial tidal disruption events. 

\section{The rate and distribution}
\label{sec:rate}
\subsection{Fraction in the Overall Population}
It is possible that a significant fraction of FRBs resides in this kind of dynamic environment.  
Up to the distance of M81, there are around 1000 globular clusters,
and at least one of them has an active FRB source, giving a lower
limit of active source number density $\rho_\mr{src}=10^{-3}$GC$^{-1}$. Adopting
a volumetric number density of GCs $\rho_{GC}=2.31$ GC Mpc$^{-3}$
\citep{2015PhRvL.115e1101R}, we can estimate the volumetric rate of the GC
FRBs from the rate of M81 bursts $R_{M81}=0.07$ hr$^{-1}$ above an energy of
$E=F\Delta\nu D^2=2.5\times 10^{33}$ erg:  
\begin{align}
    R(>E)&=\rho_{GC} \rho_\mr{src} R_{M81} E_{33}^{-\gamma} \\
    &=10^9 \mr{Gpc}^{-3}\mr{yr}^{-1} E_{33}^{-\gamma}
\end{align}
where $\gamma$ is the powerlaw index for the cumulative energy
distribution. For $\gamma=1.3$,
$R(>E_{39})=20\mr{Gpc}^{-3}\mr{yr}^{-1}$, which is much less than the
number estimated for the CHIME population \citep{2023ApJ...944..105S}. However,
for $\gamma=0.5$ as in the case seen by non-repeaters and FRB
20201124A, $R(>E_{39})=10^{6}\mr{Gpc}^{-3}\mr{yr}^{-1}$ which is more
than enough to explain the whole population. 

\subsection{The source density from the NS-MS TDE}
The NS-MS TDE can provide enough active source that explains the existence of the M81 FRB and the non-detection of similar events with the observation of GCs in the Milky way. 
 
We can estimate the number of active sources $N_s$ at a given distance d with:
\begin{align}
N_s=R_\mr{ns-ms} N_\mr{GC}\tau_c
\end{align} 
where $R_\mr{ns-ms}$ is the rate of the close encounter; $N_\mr{GC}$
is the cumulative number of globular clusters up to the distance of
$d$; and $\tau_c$ can approximate the life time. And $\tau_c$ can be
estimated with the spin down time: 
\begin{align}
    \tau_c\sim \frac{P}{\dot{P}}\sim 5\times10^5 \mr{yr}\, \pms^2 \Bpsr^{-2},
\end{align}
where the period derivative is $\dot{P}=P_\mr{rad}P^3/4\pi^2I$, and
$I\sim10^{45}$gcm$^2$ is the moment of the inertia.  

The NS-MS star collision rate per core-collapse globular cluster (CCGC) is around $R_\mr{ns-ms}\approx10^{-8}$ CCGC$^{-1}$yr$^{-1}$ calibrated to N-body simulations\citep{21Kremer}. Therefore the number density of NS with $10^{10}$G magnetic field that are spun up is $10^{-3}-10^{-2}$ Mpc$^{-3}$K $\pms^2 \Bpsr^{-2}$, where K is the fraction of NS with $\Bpsr$. And the number of active source  up to the distance of M81 with 1000 GCs (and probably 250 CCGC) will be  approximately $1 \pms^2 \Bpsr^{-2}$. The number of active source in the Milky way CCGC will be $1 \pms^2 \Bpsr^{-2}$. Therefore, it is not surprising that we didn't detect similar events with hundreds of hour observation staring at few GCs in the Milky way. 

In addition to producing FRBs, tidal disruption events (TDEs) have been
proposed as a mechanism to produce isolated neutron stars in
GCs\citep{2023arXiv230715740Y}.
These can be rapidly spinning
millisecond pulsars, or young-looking pulsars.  Isolated millsecond
pulsars could also be produced by disruption of binaries, or
potentially electron capture supernovae after merger induced collapse
(MIC) of white dwarfs.  \citet{2023MNRAS.525L..22K} promoted MIC for
the formation of young neutron stars and a young magnetar to explain
GC FRB in M81. This paper explores the alternative combination that
TDEs lead to FRBs, exploiting the observed trend of stronger giant
pulses in pulsars with large magnetic fields at the light cylinder.
Our work examines the complementary FRB energetics aspects of \citet{2023arXiv230715740Y}.  The dynamical properties of GCs allow
scaling of interaction 
rates from observables, though current modelling appear to allow both
TDE and MIC as mechanisms.
There are four young pulsars observed in the globular cluster, with
spin-down ages between $10^7$ to $10^8$ years. Assuming that this
population also comes from the close encounter of NS and the main
sequence, we expect to see 0.1-1 active source per CCGC. There are
around 150 GCs in the Milky way, and around 20\% of them are
CCGC. Therefore, we expect a couple to dozens of young pulsars in the
Milky way CCGCs. And these young pulsars are indeed preferentially
observed in the CCGC.


\subsection{Location in the GC}
A typical GC velocity dispersion $\sigma \sim 5$km/s is very small
compared to the escape velocity from the surface of a main sequence
star $v_{\rm esc}\sim$ 500km/s.  The encounter is a nearly parabolic
trajectory, with the main sequency star imparting up to twice its asymptotic
linear momentum onto the neutron star.  Taking a typical MS mass
$M\sim 0.6 M_\odot$ and neutron star mass $M_n\sim 1.4 M_\odot$, we expect
a recoil velocity comparable to $\sigma$.  This places the neutron
star onto a radial orbit going out to the half mass radius, and
objects are prefentially observed at apocentre, not at the epicentre
in the central core
region which dominates the close encounter cross section.




%



\section{Summary and prediction}
\label{sec:summary}

We have presented a scenario of FRBs powered by giant pulses of
rapid spin down
neutron stars.  Giant pulses are common in pulsars with large magnetic
fields at the light cylinder, for which these rapid spin down objects
are the extreme limit. we address the energetics challenge of FRBs with narrow beaming angles. While it is not a standard assumption in emission mechanisms, the narrow beaming angle has been observed in a bright Crab giant pulse\citep{21Bij}. 
This scenario naturally explains FRBs' presence in old populations including GCs
and early type galaxies with rate consistent with the observation.  
The short lifetime of the source accounts for the paucity in the local group.

Future precision VLBI localization of
nearby FRBs\citep{2022PASP..134i4106L}
will enable clear distinction between various scenarios
\citep{2023ApJ...944....6K}.   A recycled pulsar population tends to
trace older stellar populations\citep{1988ApJ...335..755K,1990ApJ...356..174K}. 
Other than this scenario, the dynamically-formed magnetars from white dwarf mergers are so far the only alternative that can explain the presence of the GC FRB and its observed rate\citep{Kremer2021}.
The fundamental element of this scenario hinges on the presence of
strong Doppler beaming. Plasma lensing near the source can provide
requisite spatial resolution to substantiate this hypothesis. As
demonstrated in \citet{21Bij}, lensing at a
small angle $\delta$ will undergo a Doppler boost in
frequency, with a change approximately given by $\Delta f/f\approx
\gamma \delta$, where $\gamma$ represents the Lorentz factor of the
emitting particle. Consequently, FRBs characterized by an intrinsic
narrow frequency profile may exhibit a frequency shift in the
scattering tail. This shift may manifest as an upward or downward
drift or a broadening of frequency, depending upon the geometric
configuration. Furthermore, as the time delay $\tau\propto \delta^2 $,
the frequency change against time is likely $|\Delta f/f| \propto
\gamma\sqrt{\tau}$, where $\tau$ can be approximated as the delay
corresponding to the peak of the burst. This type of frequency
variation in the scattering tail is anticipated to be 
observable in the most luminous bursts, which, as per the hypothesis,
are strongly Lorentz beamed. 

\section*{Acknowledgements}
D. Z. Li thanks E. Sterl Phinney and Bing Zhang for helpful discussion. 
Ue-Li Pen thanks Claire Ye, Vicky Kalogara, Kazumi Kashiyama and Paolo
Freire for helpful discussions, and Sujin Lee for help with plotting.
Ue-Li Pen receives support from Ontario Research Fund--research
Excellence Program (ORF-RE), Natural Sciences and Engineering Research
Council of Canada (NSERC) [funding reference number RGPIN-2019-067,
CRD 523638-18, 555585-20], Canadian Institute for Advanced Research
(CIFAR), the National Science Foundation of China (Grants
No. 11929301), Thoth Technology Inc, Alexander von Humboldt
Foundation, and the National Science and Technology Council (NSTC) of
Taiwan (111-2123-M-001-008-, and 111-2811-M-001-040-). Computations
were performed on the SOSCIP Consortium's [Blue Gene/Q, Cloud Data
Analytics, Agile and/or Large Memory System] computing
platform(s). SOSCIP is funded by the Federal Economic Development
Agency of Southern Ontario, the Province of Ontario, IBM Canada Ltd.,
Ontario Centres of Excellence, Mitacs and 15 Ontario academic member
institutions.



\bibliographystyle{mnras}
\input{example.bbl}





\bsp	
\label{lastpage}
\end{document}